\documentclass[
preprint,   
preprintnumbers,   
amsmath,amssymb,
]{revtex4-2}

\usepackage{graphicx}
\usepackage{dcolumn}
\usepackage{bm}
\usepackage{hyperref}
\usepackage{xurl}

\newcommand{\lnnavg}{\ln\langle n \rangle}
\newcommand{\navg}{\langle n \rangle}

\begin{document}

\title{An approximate formula for the entropy of the negative binomial distribution}

\author{S\'andor L\"ok\"os\footnote{sandor.lokos@cern.ch}}
\affiliation{Institute of Nuclear Physics Polish Academy of Sciences}


\begin{abstract}
Recent theoretical developments revived the interest in charged particle multiplicities and their wide-spread parametrization, the negative binomial distribution (NBD). The central observable of the studies is the Shannon entropy of the NBD. A closed form is not known, however, there are representations with special series and integrals. In this note, we will investigate one of these and give an approximate formula for the entropy that is valid up to $\sim$20\% deviation from the exact value for extreme values of the NBD parameters.
\end{abstract}

\maketitle

\section{Introduction}

Negative binomial distribution is the most common parametrization of the charged particle multiplicities measured in high energy particle reactions, in systems like $e^-p$ \cite{H1:1996ovs,H1:2020zpd}, $pp$ \cite{LHCb:2014wmv,ALICE:2015olq,ALICE:2017pcy,ATLAS:2010jvh,ATLAS:2016qux,ATLAS:2016zkp,CMS:2010qvf,UA5:1988gup} or even heavy ion collisions, e.g. \cite{PHENIX:2008psu}.

Recent studies (e.g. Refs~\cite{Kharzeev:2017qzs,Kharzeev:2021nzh,Kharzeev:2021yyf,Kutak:2023cwg,Kutak:2025syp,Hentschinski:2021aux,Kutak:2025tsx,Lokos:2025cbu}) suggest that the initial state entanglement entropy could be related to the final state Shannon entropy. Therefore the Shannon entropy of the negative binomial distribution is of great interest for the community.

Unfortunately, a simple, closed formula does not exist for the entropy of the NBD. On the other hand, there are formulae that can be evaluated numerically. Such an expression can be found in Ref.~\cite{DBLP-abs-1708-06394}, in Eq. (23) that contains a part that depends only on the parameters of the distribution and an integral that should be evaluated numerically. Following the conventional notations of the particle physics community, that partially overlap with the notation of the Wikipedia, i.e.,
\begin{align}
    p=\frac{k}{k+\navg} \qquad , \qquad (1-p) = \frac{\navg}{k+\navg}
\end{align}

Eq. (23) of Ref.~\cite{DBLP-abs-1708-06394} can be written as
\begin{align}
    S_{\rm NBD} &=-\frac{k}{p}\left[ (1-p)\ln(1-p)+p\ln(p)+(1-p)\ln(k) \right] + \nonumber \\
    &\hspace{3cm}+\int_0^1 \frac{(1-z)^{k-1}-1}{z\ln(1-z)}\left[ \left(1+\frac{(1-p)z}{p}\right)^{-k}+\frac{(1-p)kz}{p} -1 \right] = \nonumber \\
    &=-\navg\lnnavg+(\navg+k)\ln(\navg+k)-(\navg+k)\ln(k) + \nonumber \\
    &\hspace{3cm}+ \int_0^1 \frac{(1-z)^{k-1}-1}{z\ln(1-z)}\left[ \left(1+z\frac{\navg}{k}\right)^{-k}+z\navg -1 \right].
\label{eq:SNBD_full}    
\end{align}
By evaluating the integral, it can be readily seen that it is not negligible, so one cannot avoid to perform it, but that cannot be done analytically. In the next section, an approximate formula will be given for the integral, therefore for the entropy of NBD.

\section{An approximate formula for the entropy of NBD}

Let's consider the integral given in the second part of the formula in Eq.~\eqref{eq:SNBD_full}:
\begin{align}
I(k,\navg)
=
\int_0^1
\frac{(1-z)^{k-1}-1}{z\ln(1-z)}
\left[
\left(1+\frac{\navg}{k}z\right)^{-k}
+\navg z
-1
\right]dz ,
\end{align}
with \(k\geq 1\), \(\navg\geq 1\), that is expected in particle physics applications. The term that is proportional to \(\navg z\) can be evaluated analytically by using the identity
\begin{align}
\frac{(1-z)^{k-1}-1}{\ln(1-z)} = \int_0^{k-1}(1-z)^t\,dt, 
\end{align}
so the integral becomes
\begin{align}
I(k,\navg) = \navg\ln k + J(k,\navg),
\end{align}
where
\begin{align}
J(k,\navg) = \int_0^1 \frac{(1-z)^{k-1}-1}{z\ln(1-z)} \left[ \left(1+\frac{\navg}{k}z\right)^{-k} -1 \right]dz .
\end{align}
With this, we arrived to a new and a bit more simple formula than the one that is given in Ref.~\cite{DBLP-abs-1708-06394}, written as
\begin{align}
    S_{\rm NBD} =(\navg{+}k)\ln(\navg{+}k){-}\left(\navg\lnnavg{+}k\ln(k)\right){+}\int_0^1 \frac{(1{-}z)^{k-1}{-}1}{z\ln(1{-}z)}\left[ \left(1{+}z\frac{\navg}{k}\right)^{-k}{-}1\right],
\end{align}
or by rearranging the terms with logarithms
\begin{align}
    S_{\rm NBD} =-\navg\ln\left(\frac{\navg}{\navg{+}k}\right){-}k\ln\left(\frac{k}{\navg{+}k}\right){+}\int_0^1 \frac{(1{-}z)^{k-1}{-}1}{z\ln(1{-}z)}\left[ 1{-}\left(1{+}z\frac{\navg}{k}\right)^{-k}\right].
\end{align}
Unfortunately, the integral, however, is finite, cannot be evaluated explicitly further. So let's work out an approximation. A useful change of variables is
\begin{align}
u=-\ln(1-z), \qquad z=1-e^{-u}, \qquad dz=e^{-u}du .
\end{align}
Then
\begin{align}
J(k,\navg) = \int_0^\infty \frac{1-e^{-(k-1)u}}{u(e^u-1)} \left[ \left(1+\frac{\navg}{k}(1-e^{-u})\right)^{-k} -1 \right]du .
\end{align}
Now, the analytic approximation is obtained by replacing the square bracket in this representation by a single saturating exponential,
\begin{align}
\left(1+\frac{\navg}{k}(1-e^{-u})\right)^{-k}
-1
\approx
-D\left(1-e^{-\lambda u}\right).
\end{align}
This gives the analytic, approximate contribution
\begin{align}
J_{\rm approx} = -D \int_0^\infty \frac{ (1-e^{-(k-1)u})(1-e^{-\lambda u}) } {u(e^u-1)} \,du =
-D \ln \left[ \frac{\Gamma(k+\lambda)}  {\Gamma(k)\Gamma(1+\lambda)} \right].
\end{align}
The corresponding approximation for the full integral is
\begin{align}
I(k,\navg) \approx \navg\ln k - D \ln \left[ \frac{\Gamma(k+\lambda)}  {\Gamma(k)\Gamma(1+\lambda)} \right],
\end{align}
where
\begin{align}
D = 1- \left(1+\frac{\navg}{k}\right)^{-k}, \qquad \lambda = \frac{\navg}{D}.
\end{align}
Finally, putting everything together, the approximate entropy for the NBD is given as
\begin{align}
    S_{\rm NBD} \approx (\navg+k)\ln(\navg+k)-\left(\navg\lnnavg+k\ln(k)\right) - D \ln \left[ \frac{\Gamma(k+\lambda)}  {\Gamma(k)\Gamma(1+\lambda)} \right]
\label{eq:SNBD_approx}    
\end{align}
or
\begin{align}
    S_{\rm NBD} \approx-\navg\ln\left(\frac{\navg}{\navg{+}k}\right){-}k\ln\left(\frac{k}{\navg{+}k}\right){-} D \ln \left[ \frac{\Gamma(k+\lambda)}  {\Gamma(k)\Gamma(1+\lambda)} \right].
\end{align}
This approximate formula still involves the special function $\Gamma$ but that is the standard part of numerical packages like the ROOT framework \cite{ROOT_NIMA_1997}, GSL \cite{gough2009gnu} or the \texttt{scipy.special.gamma} function from Python SciPy \cite{scipy_gamma}, hence Eq.~\eqref{eq:SNBD_approx} might be implemented easier than Eq.~\eqref{eq:SNBD_full}.

\newpage

\section{Discussion}

\begin{figure}[h!]
    \centering
    \includegraphics[width=0.68\textwidth]{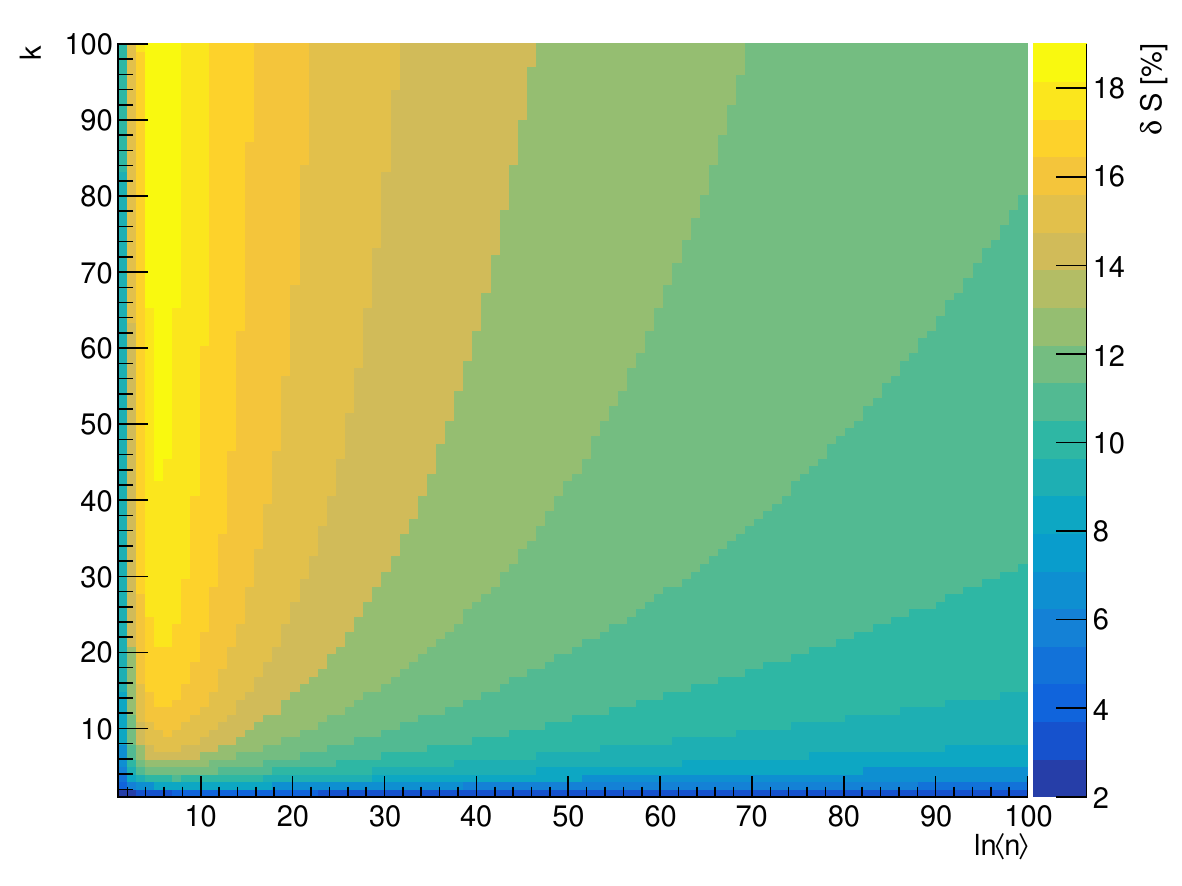}
    \caption{The $\delta S$ relative difference of the exact (numerically evaluated) entropy and the approximate entropy of the negative binomial distribution.}
    \label{fig:deltaS}
\end{figure}

The formula given in Eq.~\eqref{eq:SNBD_approx} is an approximation. The relative difference of the approximation and the ,,exact'' (numerically evaluated) entropy is calculated as
\begin{align}
    \delta S = \frac{S_{\rm approx}-S_{\rm exact}}{S_{\rm exact}}
\end{align}
and was determined to be within $\sim 20\%$ for extreme values of $\navg$ and $k$ but typically around $\sim 10\%$. A detailed scan is presented in Fig.~\ref{fig:deltaS}.

\section{Conclusion}

There is no available, closed form for entropy of the negative binomial distribution in literature, not such that we are aware of. Special series involving the Pochhammer symbols and special series can be found. In this note, an approximate formula was given that is $\sim10\%$ accurate in the typical parameter domain of the NBD and $\sim 20\%$ in the extreme, over dispersed cases. If one needs a precise determination of the entropy, the numerical implementation of the above integral cannot be avoided, but if the present precision is satisfactory, the closed form (that still involve the $\Gamma$ function) can be utilized.

\section*{References}
\bibliography{citation}

\end{document}